\documentclass[a4paper,fleqn,usenatbib]{mnras}

\usepackage[T1]{fontenc}
\usepackage{ae,aecompl}
\usepackage{hyperref}

\usepackage{graphicx}	
\usepackage{amsmath}	
\usepackage{amssymb}	
\usepackage{bm}
\usepackage{graphics}
\usepackage{xcolor}
\usepackage{newtxtext} 
\usepackage{ulem}

\title[PPN TDSL]{The first simultaneous measurement of Hubble constant and post-Newtonian parameter from Time-Delay Strong Lensing}

\author[T. Yang, S.Birrer and B. Hu]{
Tao Yang$^{1,2}$,
Simon Birrer$^{3}$
and Bin Hu$^{1}$\thanks{E-mail: bhu@bnu.edu.cn}  
\\
$^{1}$Department of Astronomy, Beijing Normal University, Beijing 100875, China\\
$^{2}$Asia Pacific Center for Theoretical Physics, Pohang 37673, Korea\\
$^{3}$Kavli Institute for Particle Astrophysics and Cosmology and Department of Physics, Stanford University, Stanford, CA 94305, USA
}

\date{Accepted XXX. Received YYY; in original form \today}

\pubyear{2020}

\begin{document}
\label{firstpage}
\pagerange{\pageref{firstpage}--\pageref{lastpage}}
\maketitle

\begin{abstract}
Strong gravitational lensing has been a powerful probe of cosmological models and gravity. To date, constraints in either domain have been obtained separately.
We propose a new methodology through which the cosmological model, specifically the Hubble constant, and post-Newtonian parameter can be simultaneously constrained. Using the time-delay cosmography from strong lensing combined with the stellar kinematics of the deflector lens, we demonstrate the Hubble constant and post-Newtonian parameter are incorporated in two distance ratios which reflect the lensing mass and dynamical mass, respectively. Through the reanalysis of the four publicly released lenses distance posteriors from the H0LiCOW collaboration,
the simultaneous constraints of Hubble constant and post-Newtonian parameter are obtained. Our results suggests no deviation from the General Relativity, $\gamma_{\texttt{PPN}}=0.87^{+0.19}_{-0.17}$ with a Hubble constant favors the local Universe value, $H_0=73.65^{+1.95}_{-2.26}$ km s$^{-1}$ Mpc$^{-1}$. Finally, we forecast the robustness of gravity tests by using the time-delay strong lensing for constraints we expect in the next few years. We find that the joint constraint from 40 lenses are able to reach the order of $7.7\%$ for the post-Newtonian parameter and $1.4\%$ for Hubble constant. 
\end{abstract}




\section{Introduction}
Over the past decade, direct measurements of the Hubble constant have achieved few percent in precision~\citep{Freedman:2017yms};
Among the conducted measurements, the Supernova H0 for the Equation of State (SH0ES) team challenge the well-believed Hubble constant ($H_0$) 
value inferred from the Planck cosmic microwave background (CMB) measurements assuming a flat $\Lambda$CDM model. 
In detail, the latest result from SH0ES is $H_0=74.03\pm1.42$ km s$^{-1}$ Mpc$^{-1}$~\citep{Riess:2019cxk}, which differs from the Planck result $H_0=67.4\pm0.5$ km s$^{-1}$ Mpc$^{-1}$ by 4.4$\sigma$~\citep{Aghanim:2018eyx}.
The Carnegie-Chicago Hubble Program (CCHP) recently also present a new and independent determination of $H_0$ parameter based on a calibration of the Tip of the Red Giant Branch (TRGB) applied to SNIa~\citep{Freedman:2019jwv,Freedman:2020dne}.
They find a value of $H_0 = 69.6 \pm 2.5$ km s$^{-1}$Mpc$^{-1}$, which is in the middle and consistent with the SH0ES and Planck values.
A different analysis and calibration of the TRGB method was performed by the SH0ES team~\citep{Yuan:2019npk,Reid:2019tiq} resutling in a slightly higher $H_0$ value.
To come to a robust conclusion, independent $H_0$ probes with accuracy better than $2\%$ are crucial~\citep{Verde:2019ivm}. 
Among the possible independent probes, the Time-Delay Strong Lensing (TDSL) measurements, such as from the $H_0$ Lenses in COSMOGRAIL's Wellspring (H0LiCOW) collaboration~\citep{Suyu:2016qxx,Bonvin:2016crt,Birrer:2018vtm,Wong:2019kwg}, are the most precise to date. 
The latest constraint from a joint analysis of six gravitationally lensed quasars with measured time delays~\citep{Wong:2019kwg} indicates for a flat $\Lambda$CDM, $H_0=73.3^{+1.7}_{-1.8}$ km s$^{-1}$ Mpc$^{-1}$, 
a $2.4\%$ precision measurement, which are in agreement with local measurements from SNIa, but in $3.1\sigma$ tension with CMB. 
The forecasts of future $40$ TDSL measurements suggest that the $H_0$ would be constrained at $\mathcal{O}(1)\%$ level~\citep{Shajib:2017omw,Yildirim:2019vlv}. 
A more optimistic forecast for dark energy studies can be found in~\citep{Shiralilou:2019div}. 
On the other hand, one of the main obstacle for the lensing mass modelling, or to determine precise $H_0$ value, is the mass-sheet degeneracy~\citep{Schneider:2013wga,Xu:2015dra}. These issues in the H0LiCOW analysis have been frequently discussed in the literature~\citep{Sonnenfeld:2017dca,Kochanek:2019ruu,Pandey:2019yic}. 
No direct evidence of bias or errors are found from a comparison of self-consistency among the individual lenses
~\citep{Millon:2019slk,Liao:2020zko}. 
Considering the fact that, both the Planck and H0LiCOW's $H_0$ values are based on General Relativity (GR) plus $\Lambda$CDM model, it inspires us to question the concordance cosmology model and investigates the modified gravity (MG).

GR has been precisely tested in various systems within our Milky Way, such as the Cassini mission within our solar system~\citep{Bertotti:2003rm}, the deflection of radio wave from the  distant compact radio sources by the sun~\citep{Shapiro:2004zz} and the energy loss via gravitational waves in the Hulse-Taylor pulsar~\citep{Taylor:1979zz}. For more references, we refer to the living review~\citep{2014LRR....17....4W}.
However, the long-range nature of gravity on the extra-galactic scale is still poorly understood.
On cosmological scale, the gravitational theory has been constrained by CMB and other observations~\citep{Ade:2015rim}. While in the non-liner regime, especially on kiloparsec (kpc) scales, GR is not fully validated with high precision. Strong gravitational lensing of galaxy can provide us with a unique opportunity to probe modifications to GR on these scales~\citep{Bolton:2006yz,Smith:2009fn,Schwab:2009nz,Cao:2017nnq,Collett:2018gpf,Yang:2018bdf,DAgostino:2020dhv}. Under the parameterized post-Newtonian (PPN) framework~\citep{Thorne:1970wv},~\citet{Collett:2018gpf} estimated $\gamma_{\texttt{PPN}}$ on the scales around $2$ kpc to be $0.97\pm0.09$ at $68\%$ confidence level by using a nearby lens, ESO 325-G004. Using the strongly lensed gravitational wave plus electromagnetic counterpart,~\citet{Yang:2018bdf} showed that the MG parameter estimation precision could be achieved at $8\%-18\%$ level. 

Though the strong lensing has been used to constrain the cosmological parameters (especially for $H_0$) and test gravity through PPN parameter ($\gamma_{\texttt{PPN}}$), the simultaneous study of the two aspects has not yet been concerned by the community.  Both the background cosmology and MG effects enter into the lens formula. On the one hand, the enhancement of gravitational force makes the lens apparently more massive. Hence, a larger Einstein radius. On the other hand, we can keep the original mass measured in GR, but reduce the distance between lens and observer. 
The covariance between these two effects should be taken into account when deriving MG constraints with strong lensing.  
With the essential cosmological information provided by the time delay, the degeneracy between gravity and background cosmology may be broken. Moreover, with the $\gamma_{\texttt{PPN}}$ as an extra parameter, the posteriors of the $H_0$ may also be changed, which provides another perspective on the aforementioned $H_0$ tension. Inspired by this motivation, we propose, for the first time, a new gravity test in TDSL. We can investigate the Hubble constant and PPN parameter simultaneously even in a single TDSL system. Through the reanalysis of four publicly released H0LiCOW's lenses~\citep{Wong:2019kwg}, the first simultaneous constrains of Hubble constant and PPN parameter are obtained. Note that, recently~\citet{Jyoti:2019pez} got the constraint, $|\gamma_{\texttt{PPN}}-1| \leq 0.2\times(\Lambda/100~{\rm kpc})$ with $\Lambda=10-200~{\rm kpc}$ by using the TDSL data. Unlike for our studies, the $H_0$ value was fixed in their inference of $\gamma_{\texttt{PPN}}$.

\section{Methodology}
In the limit of weak gravitational field, the metric of space-time is characterized by the Newtonian potential $\Psi$ and the spatial curvature potential $\Phi$,
\begin{equation}
ds^2=-\left(1+\frac{2\Psi}{c^2}\right)c^2 dt^2+a^2\left(1-\frac{2\Phi}{c^2}\right)d\vec{x}^2\,.
\label{eq:ds}
\end{equation}
The ratio $\Phi/\Psi$ is dubbed as  $\gamma_{\texttt{PPN}}$, or gravitational slip, which denotes the spatial curvature generated per unit mass. 
In the concordance model, namely, $\Lambda$CDM background evolution plus the linear structure growth following GR
\footnote{Here we follow the convention, which distinguishes DE from MG models. 
The former is focusing on the background evolution and the latter is sensitive to the structure growth pattern.}, 
$\gamma_{\texttt{PPN}}$ equals to unity or $\Psi=\Phi$.  
In this letter, we assume a constant $\gamma_{\texttt{PPN}}$ on the relevant lens galaxy scales.

The traditional idea of using the strong lensing phenomena to test gravity is via the two different mass measurements, namely, the dynamical mass obtained from the spectroscopic measurement of the stellar kinematics of the deflector galaxy, 
and the lensing/light mass inferred from the lensing image.   
The Newtonian potential ($\Psi$) is gravitational sector which responds to the motions of the non-relativistic species, such as baryonic and dark matters.
This can be easily seen from the Poisson equation. 
On the other hand, $\Phi$ is the three-dimensional spatial curvature. Its relations with the Newtonian potential are determined both by the matter components (\textit{eg.} matter species in the concordance model plus extra scalar fields) and by the gravity \textit{per se.}   
The stellar kinematics of the deflector are sensitive only to the Newtonian potential $\Psi$, whilst the lensing observables, such as lensing image, are sensitive to the Weyl potential, 
$\Psi_+=\frac{\Psi+\Phi}{2}=(\frac{1+\gamma_{\texttt{PPN}}}{2})\Psi$. Thus the comparison between dynamical mass and lensing mass is indeed the direct comparison of $\Psi$ and $\Psi_+$. 
Under the framework of PPN, the lensing mass information is encoded in the deflection angle which reads
\begin{eqnarray}
\alpha_{\texttt{PPN}}(\theta)&=&\left(\frac{1+\gamma_{\texttt{PPN}}}{2}\right)\alpha_{\texttt{GR}}(\theta) \nonumber \\
                      &=&\left(\frac{1+\gamma_{\texttt{PPN}}}{2}\right)\frac{1}{\pi}\int_{\mathbb{R}^2}d^2\theta' \frac{\Sigma(D_{\rm d}\theta')}{\Sigma_{\rm cr}}
                      \frac{\theta-\theta'}{|\theta-\theta'|^2}\,,
\label{eq:alpha}
\end{eqnarray}
where $\theta$ is the image position, $\Sigma$ is the surface mass density
and $\Sigma_{\rm cr}=\frac{c^2}{4\pi G}\frac{D_{\rm s}}{D_{\rm d}D_{\rm ds}}$ is the critical surface mass density which depends on the angular diameter distances of source and lens, hence background cosmology.

A constant $\gamma_{\texttt{PPN}}$ rescales the Newtonian potential $\Psi$, effective lensing potential $\psi$ (the integral of the Weyl potential along the line-of-sight) as well as convergence field $\kappa$ in GR. 
In details, they read $\Psi_+=(\frac{1+\gamma_{\texttt{PPN}}}{2})\Psi$, $\psi_+=(\frac{1+\gamma_{\texttt{PPN}}}{2})\psi$ and $\kappa'=(\frac{1+\gamma_{\texttt{PPN}}}{2})\kappa$. 
Eq.~(\ref{eq:alpha}) tells us, that the MG effect and cosmological distance are degenerated.  
For instance, $\gamma_{\texttt{PPN}}>1$ can be interpreted as the enhancement of gravitational force. 
The deflection angle in this case is larger than the one in GR. 
On the other hand, we can keep the gravity unmodified, but change the corresponding distances.
These two effects are highly degenerated. It is one of the major limitation of using strong lensing events to test gravity.

To overcome this problem, additional data, either the cosmological or the gravitational one, are needed.  
In this letter, we propose that the TDSL data are able to break the degeneracy alone. 
The time delay between image A and B is
$\Delta t_{\rm AB}=\frac{D_{\Delta_t}}{c}\left[\phi(\theta_{\rm A},\beta)-\phi(\theta_{\rm B},\beta)\right]=\frac{D_{\Delta_t}}{c}\Delta\phi_{\rm AB}(\xi_{\rm lens})$~\citep{Shapiro:1964uw}. Here $\phi(\theta,\beta)=\left[\frac{(\theta-\beta)^2}{2}-\psi(\theta)\right]$ is the Fermat potential, $\beta$ is the source position, $\xi_{\rm lens}$ is the lens model parameter.
$D_{\Delta t}=(1+z_{\rm d})\frac{D_{\rm d}D_{\rm s}}{D_{\rm ds}}$ is the time-delay distance which is inverse proportional to $H_0$.  
These formulas are valid in all background cosmologies and gravity models, as long as the latter are metric theories. 
In the time-delay equation, the Fermat potential difference $\Delta\phi_{\rm AB}(\xi_{\rm lens})$ is reconstructed from the lensing image. However, as we mentioned previously, under the PPN framework, the inferred mass parameters are rescaled by a factor of $(1+\gamma_{\texttt{PPN}})/2$. Hence, we denote the actually inferred lens model parameters in the Fermat potential as $\xi'_{\rm lens}$. In this case, the time-delay distance can be written as
\begin{equation}
D_{\Delta t}=(1+z_{\rm d})\frac{D_{\rm d}D_{\rm s}}{D_{\rm ds}}=\frac{c\Delta t_{\rm AB}}{\Delta\phi_{\rm AB}(\xi'_{\rm lens})}\,.
\label{eq:ddt}
\end{equation}
The left-hand side of Eq.~(\ref{eq:ddt}) is the first distance ratio we need. It is calculated from both the measurement of time delay and the Fermat potential reconstructed with parameter $\xi'_{\rm lens}$. The only difference of this equation under the PPN framework is that the inferred lens model parameters are $\xi'_{\rm lens}$ but not the original $\xi_{\texttt{lens}}$ under GR.

To incorporate the stellar kinematics information, we follow the parametric method used by H0LiCOW collaboration~\citep{Suyu:2009by,Birrer:2015fsm,Birrer:2018vtm} (and references therein).
The radial velocity dispersion $\sigma_r$ is modelled via the anisotropic Jeans equation
\begin{equation}
\frac{\partial(\rho_*\sigma_r^2)}{\partial r}+\frac{2\beta_{\texttt{ani}}(r)\rho_*\sigma_r^2}{r}=-\rho_*\frac{\partial\Psi}{\partial r}\,,
\label{eq:jeans}
\end{equation}
where $\beta_{\texttt{ani}}(r)\equiv 1-\frac{\sigma_t^2}{\sigma_r^2}$ is the stellar distribution anisotropy. $\sigma_t$ is the tangential dispersion.  $\rho_*$ is the luminosity distribution of the lens. The luminosity-weighted projected velocity dispersion $\sigma_s$ is given by~\cite{Suyu:2009by}, $I(R)\sigma_s^2=2\int_R^\infty \left(1-\beta_{\texttt{ani}}(r)\frac{R^2}{r^2}\right)\frac{\rho_*\sigma_r^2rdr}{\sqrt{r^2-R^2}}$, 
where $R$ is the projected radius and $I(R)$ the projected light distribution.  Finally, the luminosity-weighted line-of-sight velocity dispersion within an aperture, $\mathcal{A}$, is then $\sigma_v^2=\frac{\int_\mathcal{A}[I(R)\sigma_s^2*\mathcal{P}]dA}{\int_\mathcal{A}[I(R)*\mathcal{P}]dA}$.
Furthermore, we can write $\sigma_v$ in terms of $D_s/D_{ds}c^2J(\xi_{\texttt{lens}},\xi_{\texttt{light}},\beta_{\texttt{ani}})$~\citep{Birrer:2018vtm}, where 
$\xi_{\texttt{light}}$ denote for the light model parameters.  
The function $J$ captures all the ingredients for computing the velocity dispersion.

The distance ratio from the stellar kinematics (dynamical mass) is~\citep{Birrer:2018vtm}
\begin{equation}
\frac{D_{\rm s}}{D_{\rm ds}}=\frac{\sigma_v^2}{c^2J(\xi_{\texttt{lens}},\xi_{\texttt{light}},\beta_{\texttt{ani}})}\;.
\label{eq:gr_dsdds}
\end{equation}
Since stellar dynamics are determined only by the Newtonian potential $\Psi$, this distance ratio shall not be influenced by the PPN parameter. 
The lens model parameter in $J$ is the ``unrescaled'' $\xi_{\texttt{lens}}$.
If we replace $\xi_{\texttt{lens}}$ with $\xi'_{\texttt{lens}}$, the resulted distance ratio shall also be rescaled, correspondingly
\begin{equation}
\frac{2}{1+\gamma_{\texttt{PPN}}}\frac{D_{\rm s}}{D_{\rm ds}}=\frac{\sigma_v^2}{c^2J(\xi'_{\texttt{lens}},\xi_{\texttt{light}},\beta_{\texttt{ani}})}\,.
\label{eq:dsdds1}
\end{equation}
Furthermore, we can define $D_{\rm d}'=\frac{1+\gamma_{\texttt{PPN}}}{2}D_{\rm d}$. 
By combining Eqs.~(\ref{eq:ddt}) and (\ref{eq:dsdds1}), we get 
\begin{eqnarray}
D_{\rm d}'=\frac{1}{1+z_{\rm d}}\frac{c\Delta t_{\rm AB}}{\Delta \phi_{\rm AB}(\xi'_{\texttt{lens}})}\frac{c^2J(\xi'_{\texttt{lens}},\xi_{\texttt{light}},\beta_{\texttt{ani}})}{\sigma_v^2}\,.
\label{eq:ddp}
\end{eqnarray}
This is the second distance we need.

In summary, combining the three measurements, namely optical lensing image, deflector spectroscopies 
as well as time delays,
we can get two distances ($D_{\Delta t}, D_{\rm d}'$), simultaneously, from Eqs.~(\ref{eq:ddt}) and (\ref{eq:ddp}).
The second distance ($D_{\rm d}'$) carries the information of alternative theories to GR. 
We can directly use the public posteriors of $D_{\Delta t}$ and $D_{\rm d}$ and rescale $D_{\rm d}$ to $D'_{\rm d}$ according to Eq. (\ref{eq:ddp}).  

After introducing the methodology, some issues in the realistic lensing analysis should be concerned. 
First of all, the nearby perturbering galaxies can induce extra external shears~\citep{Wong:2016dpo,Birrer:2018vtm}. 
Secondly,
the masses distributed along the line of sight can also alter the inferred distance ratios as this
causes extra focusing and defocusing of the light rays and can affect the observed time delays~\citep{Seljak:1994wa}.
On large scales, the external convergence ($\kappa_{\texttt{ext}}$) is generally of order few per cent. 
Their effects on the resulting distance estimation can be parametrized as $D_{\Delta t}=D_{\Delta t}^{\texttt{model}}/(1-\kappa_{\texttt{ext}})$~\citep{Keeton:2002ug,McCully:2013fga,Suyu:2009by}, 
where $D_{\Delta t}^{\texttt{model}}$ is the time-delay distance without the external convergence. 
Generally, $\kappa_{\texttt{ext}}$ cannot be constrained from the lens model due to the mass-sheet degeneracy~\citep{Saha:2000kn}.  
To break this degeneracy, one have to study the environment and estimate the mass distribution along the line-of-sight via simulation~\citep{Fassnacht:2005uc,Momcheva:2005ex,Williams:2005ew,Wong:2010xk}.
We do not include the MG effect in the estimation of $\kappa_{\texttt{ext}}$ in this work. We expect the impact on $\kappa_{\texttt{ext}}$ to be subdominant on our derived constraints. Our focus is on the gravity test on the deflector galaxy scales.
And, we model the MG effect via a simple constant $\gamma_{\texttt{PPN}}$ parameter. 
This parametrization can not be extended to the cosmological scales.  
Hence, we make use of the standard analysis of the external convergence/shear in H0LiCOW collaboration~\citep{Birrer:2018vtm,Wong:2019kwg}.

\section{Results}
\begin{figure*}
\includegraphics[width=0.8\textwidth]{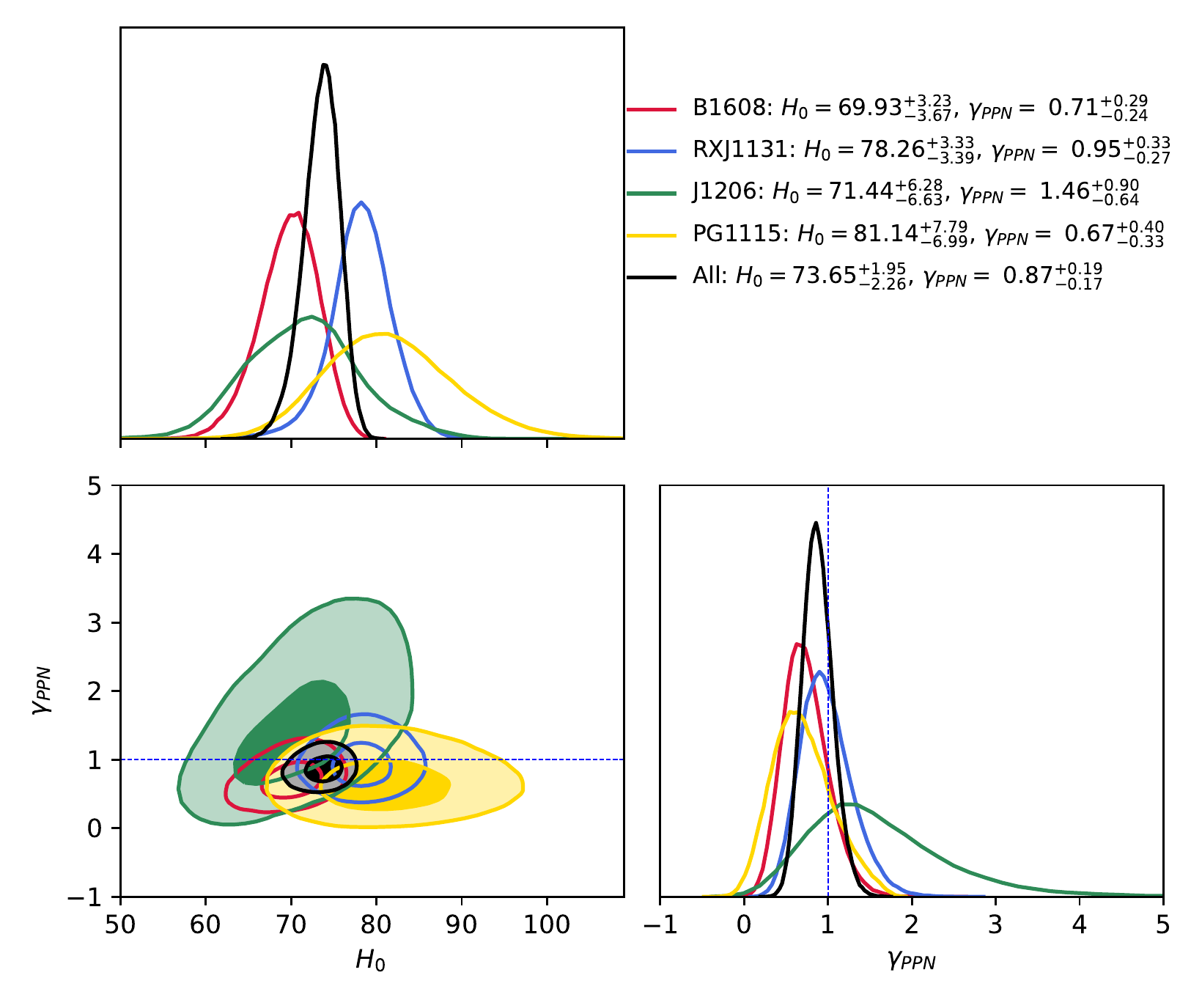}
\caption{The constraints of $H_0$ and $\gamma_{\texttt{PPN}}$ from four of the H0LiCOW lenses. The dashed line is $\gamma_{\texttt{PPN}}=1$ predicted by GR. Priors are: $H_0$ uniform in [0, 150] km s$^{-1}$ Mpc$^{-1}$ and  $\Omega_m$ uniform in [0.05, 0.5]. All the lenses show to be consistent with GR within 1 $\sigma$ confidence level.}
\label{fig:result_h0licow}
\end{figure*}

H0LiCOW collaboration has analyzed six strong lensing systems, four of which ({\it B1608, RXJ1131, J1206, PG1115}) have both $D_{\Delta t}$ and $D_{\rm d}$ measurements. Details of the analyses can be found in H0LiCOW's papers~\citep{Suyu:2009by,Jee:2019hah,Chen:2019ejq,Suyu:2013kha,Huber:2019ljb,Wong:2016dpo,Birrer:2018vtm,Rusu:2019xrq,Wong:2019kwg}. Having the public posteriors of $D_{\Delta t}$ and $D_{\rm d}$\footnote{\url{http://www.h0licow.org}} for these four lenses, we incorporate the PPN parameter $\gamma_{\texttt{PPN}}$ into these two distances. We modify the Python notebook by~\citet{martin_millon_2020_3633035} to infer the relevant parameters. Assuming flat $\Lambda$CDM cosmology, and adopting Markov Chain Monte Carlo (MCMC) from the Python package {\sc emcee}~\citep{ForemanMackey:2012ig}, the constraints of Hubble constant $H_0$, matter density $\Omega_m$\footnote{$\Omega_m$ is not well constrained and we do not show it in the results.} and PPN parameter $\gamma_{\texttt{PPN}}$ are obtained. The results are shown in Fig.~\ref{fig:result_h0licow}.  
We find that the PPN parameter values inferred from the four individual lenses are consistent with General Relativity. The joint constraint gives $\gamma_{\texttt{PPN}}=0.87^{+0.19}_{-0.17}$, $H_0=73.65^{+1.95}_{-2.26}$ km s$^{-1}$ Mpc$^{-1}$.\footnote{The constraints of $H_0$ and $\gamma_{\texttt{PPN}}$ for each individual lenses can be found in Fig.~\ref{fig:result_h0licow}.}
The varying orientation of the ellipses reflect the different weight of the kinematics information in the H0LiCOW analysis on their $H_0$ inference. The kinematic information in the H0LiCOW analysis is imprinted in the inferred angular diameter distance to the lens, $D'_{\rm d}=D_{\rm d}(1+\gamma_{\texttt{PPN}})/2$, which is fully degenerate and thus reflect degeneracies between $H_0$ and $\gamma_{\texttt{PPN}}$ in our analysis. \cite{Millon:2019slk} showed that the dispersion measurements do not play a significant role in the $H_0$ estimation of the H0LiCOW analysis, except for {\it J1206}.

Moreover, we forecast the future constraints through the simulated TDSL. The simulation is based on the Python package {\textsc lenstronomy}\footnote{\url{https://github.com/sibirrer/lenstronomy}}~\citep{Birrer:2018xgm}. In the simulation, we assume the fiducial model as GR and $\Lambda$CDM cosmology with $H_0=70$ km s$^{-1}$ Mpc$^{-1}$ and $\Omega_m=0.3$  . The lens model we choose is a singular elliptical power-law model with an external shear. A source position is fixed so that multiple images are produced. From Eqs.~(\ref{eq:ddt}) and (\ref{eq:ddp}), the uncertainties of the two distance posteriors come from the reconstructions of  $\Delta \phi$ and $J$, the measurements of the time delay $\Delta t$ and velocity dispersion $\sigma_v$, and finally $\kappa_{\texttt{ext}}$ along the line-of-sight. We assign the slope of the power-law lens model $\gamma'$ to represent the parameter which propagates the error of lens model reconstructions to the uncertainties of $\Delta \phi$ and $J$. 

\begin{figure*}
\includegraphics[width=0.8\textwidth]{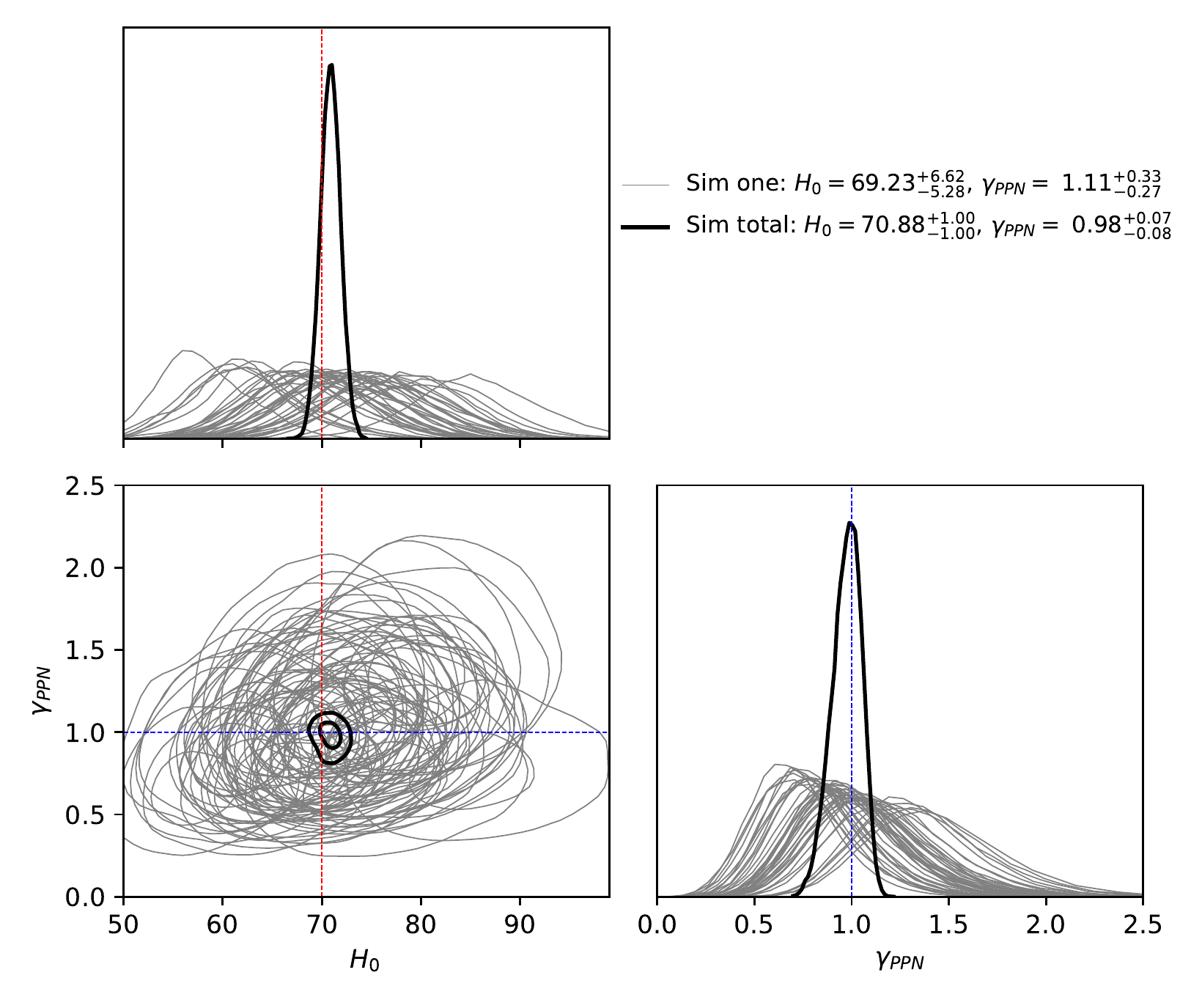} 
\caption{The forecast of 40 lenses based on the error estimates presented in Table~\ref{tab:uncertainty}. ``Sim one'' represents a typical result for one of the 40 lenses, ``Sim total'' is the joint constraint from the total 40 lenses. The dashed line represents the fiducial value we set for $H_0$ and $\gamma_{\texttt{PPN}}$.}
\label{fig:result_sim}
\end{figure*}

The future TDSL cosmography has been forecasted by~\citet{Shajib:2017omw} with spatially resolved kinematics and by~\citet{Yildirim:2019vlv} using high signal-to-noise integral field unit observations from the next generation telescopes. 
The former predicts roughly $6\%$ and $10\%$ errors on $D_{\Delta t}$ and $D_{\rm d}$, while the later claims that, for {\it RXJ1131}-like system, $D_{\Delta t}$ and $D_{\rm d}$  can be even constrained to $2.3\%$ and $1.8\%$, respectively.
In this letter, we simulate 40 strong lensing systems with redshift distribution according to~\cite{Shajib:2017omw}.
We generate the redshift distribution for our forecast as follow: We fit a Gaussian distribution to the redshift distribution of deflectors of the current 6 lenses from H0LiCOW. Then we sample the remaining 34 lens redshifts from this fitted Gaussian distribution. The same procedure is applied for the source redshift distribution.
The parameter uncertainties we adopted for the simulation are summarized in Table~\ref{tab:uncertainty}. 
The inferred precision on $D_{\Delta t}$ and $D_d$ are 8\% and 14\% respectively. In this conservative scenario, the errors of $D_{\Delta t}$ and $D_d$ are worse by $2\sim4$\% compared to~\cite{Shajib:2017omw}. The forecast results are shown in Fig.~\ref{fig:result_sim}.
As noted in~\cite{Collett:2018gpf}, for a galaxy-galaxy lensing system like ESO 325-G004, the systematic velocity dispersion bias is the limiting factor for gravity tests with current data. If all lenses have a common $x\%$ systematic velocity dispersion offset, this would lead to a $2x\%$ systematic error in $\gamma_{\texttt{PPN}}$. As shown in Tab.~\ref{tab:uncertainty}, we conservatively took $5\%$ statistical error in velocity dispersion. However, we did not consider the aforementioned systematic error in the forecast.

\begin{table*}
\caption{Parameters setup for the simulation. $\gamma'$ is the slope of the lens mass profile which we set to denote the lens model. ``astrometry'' is the uncertainty for measuring the image position. The reconstruction errors of the two distances from the setup errors are also shown.} 
\begin{tabular}{c c c c c | c c}
\hline\hline
\multicolumn{5}{c |}{Setup errors} & \multicolumn{2}{c}{Inferred distances errors} \\
\hline
$\delta\gamma'$ ~&~ astrometry (arcsec) ~&~ $\delta\Delta t/\Delta t$ ~&~ $\delta\kappa_{\texttt{ext}}$ ~&~ $\delta\sigma_v/\sigma_v$ ~&~ 
$D_{\Delta t}$ ~&~ $D_{\rm d}$ \\
\hline
0.02  ~&~ 0.005 ~&~ 2\% ~&~ 0.03 ~&~ 5\% ~&~ $8\%$ ~&~ $14\%$ \\
\hline
\end{tabular}
\label{tab:uncertainty}
\end{table*}

\noindent
\textit{Conclusions and discussions.}
Strong gravitational lensing provides us a powerful probe for cosmology as well as the gravity theory. These two aspects have been studied separately by the community in the literatures. In this letter we incorporate the comparison of lensing mass and dynamic mass for the test of GR into the time-delay cosmography. The simultaneous constraints of $H_0$ and $\gamma_{\texttt{PPN}}$ for four lenses of the current H0LiCOW analysis are obtained.
In our analysis, we explicitly assume $\gamma_{\texttt{PPN}}$ to be a constant across the Einstein radius of each lens galaxies. To be specific, they are $5.46$ kpc (for B1608), $6.42$ kpc (for RXJ1131), $9.30$ kpc (for J1206), and $4.26$ kpc (for PG1115), respectively.
We find that the PPN parameter values inferred from all these four lenses are consistent with General Relativity. The joint constraint gives $\gamma_{\texttt{PPN}}=0.87^{+0.19}_{-0.17}$, $H_0=73.65^{+1.95}_{-2.26}$ km s$^{-1}$ Mpc$^{-1}$. The Hubble constant value is not significantly different from the one assuming GR.

Recent TDCOSMO's paper~\citep{Millon:2019slk} investigates the possible systematic errors and biases introduced from such as stellar kinematics and lens model. No significant trends indicative of biases are reported. In this letter, we consistently check the relationship between $D_{\Delta t}$ and $D_{\rm d}$ as a sign of MG effect. We may also treat our study as a test of the compatibility between the stellar kinematics and the adopted lens model under GR. Our results also suggest there is no significant inconsistency from this perspective.

For the future forecast, we use a realistic error estimation and simulate 40 lenses based on the existed lenses redshift distribution. The robustness of the constraints on $H_0$ and $\gamma_{\texttt{PPN}}$ is determined by the errors of the two distances, $D_{\Delta t}$ and $D_d$. In our simulation, one typical TDSL system gives the uncertainties for $D_{\Delta t}$ and $D_{\rm d}$ at the order of $8\%$ and $14\%$. It can constrain the $H_0$ and $\gamma_{\texttt{PPN}}$ at the order of $8\%$ and $27\%$. 
The 40 TDSL events are jointly able to reach the order of $1.4\%$ and $7.7\%$ in $H_0$ and $\gamma_{\texttt{PPN}}$, respectively. If we take a more optimistic estimation~\cite{Yildirim:2019vlv}, the future errors on $D_{\Delta t}$ and $D_{\rm d}$ can even decrease dramatically to the order of $2.3\%$ and $1.8\%$, which indicates a more promising future of TDSL as the probe of cosmology and gravity theory.

\noindent
\textit{Acknowledgements.}
BH thank Sherry Suyu for helpful discussion. 
TY is supported by an appointment to the YST Program at the APCTP through the Science and Technology Promotion Fund and Lottery Fund of the Korean Government, and the Korean Local Governments - Gyeongsangbuk-do Province and Pohang City. SB thanks the H0LiCOW and TDCOSMO team for useful discussion and valuable input.
BH is supported by the National Natural Science Foundation of China Grants No. 11973016, No. 11690023 and No. 11653003.



\bibliographystyle{mnras}
\bibliography{ppn_lens_refs} 







\bsp	
\label{lastpage}
\end{document}